\documentclass[sn-mathphys]{sn-jnl}

\usepackage{array}
\usepackage{colortbl}
\usepackage{lineno}
\usepackage{caption}
\usepackage{subcaption}
\usepackage{graphicx}
\usepackage{multirow}        

\makeatletter
\newenvironment{figurehere}
{
    \def\@captype{figure}
}

{}
\makeatletter

\usepackage{makecell}
\usepackage{pifont}
\usepackage[version=4]{mhchem}
\usepackage{float}
\usepackage{multicol}

\usepackage{geometry}

\usepackage{xr}
\jyear{2023}%

\theoremstyle{thmstyleone}%
\theoremstyle{thmstyletwo}%

\theoremstyle{thmstylethree}%

\begin{document}



\title[MOCI]{Multi-dimensional optical imaging on a chip}

\author*[]{\fnm{Liheng} \sur{Bian}}\email{bian@bit.edu.cn}
\equalcont{These authors contributed equally to this work.}

\author[]{\fnm{Zhen} \sur{Wang}}
\equalcont{These authors contributed equally to this work.}

\author[]{\fnm{Pengming} \sur{Peng}}
\author[]{\fnm{Zhengyi} \sur{Zhao}}
\author[]{\fnm{Rong} \sur{Yan}}
\author[]{\fnm{Hanwen} \sur{Xu}}
\author*[]{\fnm{Jun} \sur{Zhang}}\email{zhjun@bit.edu.cn}

\affil[]{State Key Laboratory of CNS/ATM \& MIIT Key Laboratory of Complex-field Intelligent Sensing, Beijing Institute of Technology, Beijing 100081, China}

\abstract
%
%
Light inherently consists of multiple dimensions beyond intensity, including spectrum, polarization, etc. The coupling among these high-dimensional optical features provides a compressive characterization of intrinsic material properties. Because multiple optical dimensions are intrinsically coupled rather than independent, analyzing their inter-relationships and achieving their simultaneous acquisition is essential.
Despite the existing optical techniques to obtain different-dimensional data with cumbersome systems, joint acquisition of multi-dimensional optical information on a chip is still a serious challenge, limited by intensity-only photoelectric detection, single-dimensional optical elements, and finite bandwidth. 
In this work, we report a multi-dimensional on-chip optical imaging (MOCI) architecture, which is functionally composed of three layers, including a multi-dimensional encoding layer to simultaneously encode different dimensions of incident light, an image acquisition layer to collect coupled intensity data, and a computational reconstruction layer to recover multi-dimensional images from a single frame of coupled measurement. Following the MOCI architecture, we for the first time fabricated a real-time (74 FPS) on-chip polarization-hyperspectral imaging (PHI) sensor, with 2048$\times$2448 pixels at 61 spectral channels covering the VIS-NIR range and 4 polarization states. We applied the PHI sensor for simultaneously resolving hyperspectral and polarization information of complex scenes, and for the first time demonstrated new applications including hyperspectral 3D modeling with normal and height maps, and hyperspectral sensing against strong reflection and glare.
The reported approach enables efficient multi-dimensional optical imaging on an integrated, low-cost, and compact chip. We anticipate the technique can enhance the understanding of the correlations among multiple optical dimensions, and open new venues for next-generation image sensors and machine intelligence.


%

\keywords{multi-dimensional imaging, hyperspectral imaging, polarization imaging, on-chip image sensor}

\maketitle

\section{Introduction}\label{sec1}

	Traditional optical imaging techniques primarily capture light intensity, while light contains richer details beyond intensity, including but not limited to spectral and polarization information \cite{rubin2019matrix,song2023helical,fan2024dispersion}. 
    These high-dimensional light information provide more compressive representations of objects' physical, chemical, and morphological properties in the real world, offering enhanced detection capabilities for next-generation machine intelligence \cite{sun2013large,liodakis2023optical,lu2024bright,cameron2024adaptive}.
	For example, hyperspectral imaging extends spatial intensity to spatio-spectral data cubes that contain hundreds or even thousands of spectral channels, enabling precise material identification \cite{ma2022intelligent}. With these advantages, hyperspectral imaging has been widely applied in various fields including remote sensing, medical diagnostics, agricultural monitoring, and machine vision \cite{backman2000detection,dremin2021skin,liu2025adaptive}.
	Similarly, polarization imaging enhances spatial imaging with additional polarization information that characterizes the vector distribution of structured light beams, revealing target shape, texture, stress, and so on \cite{spottiswoode1874polarisation}. This technique has crucial applications in fields such as biomedical imaging, remote sensing, and industrial inspection \cite{he2021polarisation,liang2019reconfigurable}.

    As current photodetectors are typically limited to intensity-only acquisition due to speed-constrained photoelectric conversion, high-dimensional imaging remains challenging and requires additional optical and mechanical elements to separately capture high-dimensional information \cite{yuan2023geometric,fan2024dispersion,oripov2023superconducting}.
	For example, traditional spectral imaging systems commonly employ grating, prisms, or narrowband filters to acquire spectral cubes through spatial or spectral scanning \cite{green1998imaging}. Likewise, conventional polarization imaging techniques typically rely on polarizers, polarizing beam splitters, or liquid crystal variable retarders to obtain polarization information \cite{fan2023disordered,wei2023geometric}. These systems usually involve complex mechanical structures and optical components, which limit the spatial, spectral, and temporal resolution, and make it challenging to achieve system integration and miniaturization \cite{hu2024metasurface}. 
    To address these challenges, innovations in materials science and nanofabrication technologies have driven the miniaturization and on-chip integration of high-dimensional imaging devices, with metasurfaces \cite{yesilkoy2019ultrasensitive,rubin2019matrix,fan2023disordered,zhang2024real}, Fabry-Pérot filter arrays \cite{yako2023video}, broadband spectral modulation dyes \cite{bian2024broadband}, aluminum nanowires \cite{gruev2010ccd}, color thin films \cite{zhang2023handheld}, liquid crystal materials being widely utilized.
    

    Despite significant advancements in high-dimensional information acquisition, jointly obtaining multiple high-dimensional light information simultaneously on a single chip remains a big challenge \cite{rogers2021universal,zhang2024real}. These multi-dimensional data are typically interdependent on each other, with characteristics such as polarization and phase of an object are closely related to the wavelength of spectrum \cite{baek2020image}. This is because different wavelengths exhibit varying scattering or reflection properties, leading to discrepancies in multi-dimensional characteristics.
    In this regard, by digging deep into the strengths of multi-dimensional optical imaging, the sensing capabilities can be significantly enhanced to advance our understanding of light properties and material characteristics \cite{altaqui2021mantis,tang2024chip}. 
    For example, benefiting from light-field control of metasurfaces, they have been applied in joint spectral and polarization imaging \cite{altaqui2021mantis,zhang2024real,jiang2024metasurface}. However, such methods face several challenges, including strong dependence on input light angle, high fabrication costs, difficulties in large-scale processing, and limitations in spatial resolution, which restrict their flexibility and practical applicability \cite{moitra2015large,he2018high}. 
    With the advancement of micro-electromechanical systems (MEMS) \cite{saccone2022additive,zhang2022large}, the integration of MEMS with two-dimensional materials or photonic crystal slabs has been reported for multi-dimensional sensing \cite{tang2024chip,tang2025adaptive}. However, these methods encounter complex electrical and mechanical components with limited stability, detection range, and field of view. Additionally, optical thin films with spatial and frequency dispersion have been applied in spectral and polarization encoding. However, this approach requires complex components for imaging, making it challenging to achieve high spatial resolution on-chip \cite{fan2024dispersion}. 

\begin{figurehere}
    \centering
    \includegraphics[width=1\linewidth]{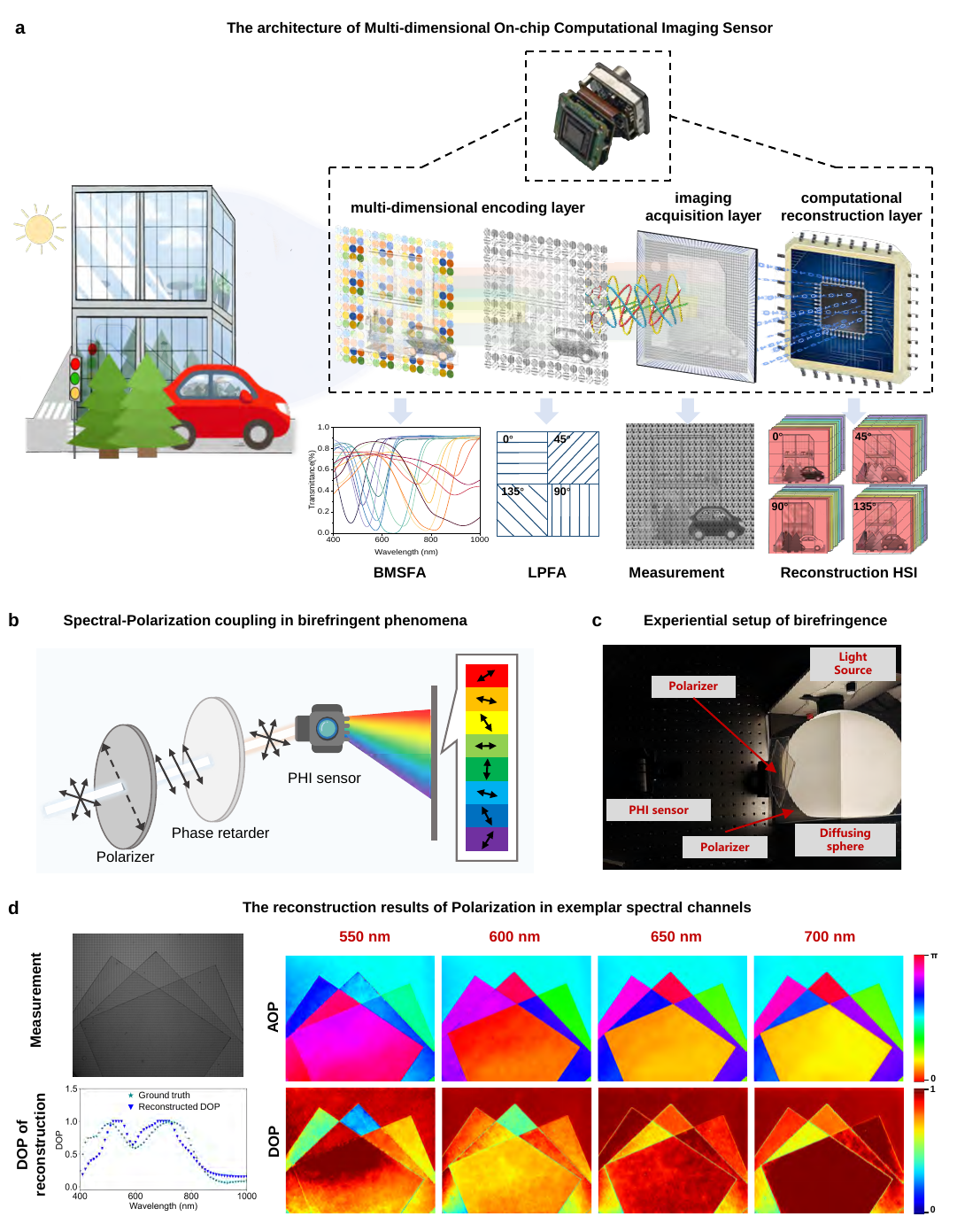}
    \caption{\label{fig_principle} {\textbf{The principle and performance of MOCI architecture.}}
        {\textbf{a.}} The MOCI architecture consists of three functional layers: a multi-dimensional encoding layer, an imaging acquisition layer, and a computational reconstruction layer. The encoding layer comprises multiple sublayers that modulate the incident light along specific dimensions. For the polarization-hyperspectral imaging (PHI) sensor, the encoding layer consists of two sublayers: a broadband multispectral filter array (BMSFA) and a linear polarization filter array (LPFA), which modulate light along spectral and polarization dimensions, respectively.
        The BMSFA is fabricated using photolithography and consists of a cyclic 4 $\times$ 4 arrangement of broadband organic spectral modulation materials. The LPFA is processed through multilayer dielectric coating and consists of 2 $\times$ 2 arrangement of linear polarizers at specific angles.
        {\textbf{b.}} Spectral-Polarization coupling in birefringent phenomena. 
        {\textbf{c.}} Experiential setup of birefringence. The light source produces uniform illumination using a diffusing sphere, then passes through a polarizer to generate polarized light, which subsequently passes through a birefringent plastic sheet before being captured by the PHI sensor. 
        {\textbf{d.}} The measurements acquired by the PHI sensor are reconstructed to reveal the degree of polarization (DOP) and angle of polarization (AOP) at different wavelengths. The reconstructed DOP of different wavelengths is compared. The reconstruction results of Polarization in exemplar spectral channels. 
        \\}		
\end{figurehere}

	To address the above challenges, this work reports a Multi-dimensional On-Chip optical Imaging (MOCI) architecture that can jointly acquire multiple high-dimensional light information. As shown in Fig. \ref{fig_principle}, the MOCI architecture is functionally composed of three layers, including the multi-dimensional encoding layer, the image acquisition layer, and the computational reconstruction layer. The multi-dimensional encoding layer contains multiple modulation sub-layers, with each one modulating a specific dimension (e.g. spectrum, polarization, phase, etc.) of incident light. Through this multi-sublayer structure, multiple dimensions of light information can be simultaneously encoded at light speed, which is then coupled into the intensity dimension and captured by the subsequent image acquisition layer. Finally, the underlying multi-dimensional optical images are reconstructed from the coupled acquired image in the computational reconstruction layer, using compressive sensing or deep learning algorithms. Such an architecture maintains the same fundamental performance of the underlying acquisition layer including spatial resolution and frame rate, while each frame can furthermore produce multiple high-dimensional information in a lightweight on-chip manner. Besides, the encoding-decoding principle compresses high-throughput information into a small number of acquired data, which can effectively decrease data amount under a limited bandwidth.

	Building on this architecture, we for the first time fabricated an on-chip polarization-hyperspectral imaging (PHI) sensor. This sensor integrates micro-nano fabricated broadband multispectral filter arrays (BMSFA) and linear polarization filter arrays (LPFA) together, enabling simultaneous control of incident light in the spectral and polarization dimensions. Specifically, the incident light first passes through the BMSFA for spectral modulation, and then through the LPFA for polarization encoding. The coupled multi-dimensional information is subsequently captured by a CMOS image chip. Finally, the original polarization-hyperspectral images are reconstructed from the acquired single intensity image using a joint-reconstruction neural network. The spectral range of the PHI sensor is from 400 to 1000 nm, with an average spectral resolution of 5 nm.
	The spatial resolution of hyperspectral images (HSIs) is 2048 $\times$ 2448, with a frame rate of 74 FPS and 61 channels. In addition, each pixel can simultaneously produce signals across four linear polarization angles to retrieve the polarization information for each spectral channel, with the linear polarization extinction ratio being as high as 5 orders of magnitude. To sum, the data throughput contains 2048 $\times$ 2448 pixels $\times$ 61 channels $\times$ 4 angles $\times$ 74 FPS.

	To demonstrate the practicality and potential applications of the fabricated PHI sensor, we conducted a series of experiments including hyperspectral separation of reflected and transmitted scenes, high-dynamic-range hyperspectral imaging with glare removal, and hyperspectral 3D modeling. Specifically, when observing objects through glass, the imaging quality is typically compromised by reflections from the surrounding scene. Using our PHI sensor, we successfully separated and reconstructed the spectra of reflected and transmitted light from one snapshot. In the interference of glare light, which typically degrades spectral imaging quality, the PHI sensor employed the liner polarization property and achieved high-dynamic-range hyperspectral imaging. Furthermore, due to the three-dimensional structural information embedded in polarization data, the PHI sensor concurrently captures the three-dimensional shape, surface normals, and spectral details of objects. To sum up, the experiments validate that the PHI sensor can not only tackle challenging lighting conditions such as reflections and glare, but also facilitate the acquisition of multi-dimensional features using a single image chip. We anticipate that the reported MOCI architecture and PHI sensor can advance the next-generation intelligent image sensor beyond CMOS or CCD in the last decades, and open new venues for future machine intelligence with multi-dimensional imaging and sensing.

\section{Fabrication and performance of the PHI sensor}\label{sec2}

The MOCI architecture employs a hierarchical design to encode, acquire, and resolve multi-dimensional information. The multi-dimensional encoding layer contains multiple modulation sub-layers, with each modulating a specific optical dimension. The image acquisition layer uses a two-dimensional image chip to capture the coupled multi-dimensional optical information as intensity measurements. Finally, the computational reconstruction layer is employed to restore the underlying multi-dimensional images from the coupled measurements using compressive sensing or deep learning algorithms.

To demonstrate the MOCI architecture, we fabricated an on-chip Polarization-Hyperspectral Imaging (PHI) sensor that can simultaneously capture high-dimensional hyperspectral and polarization images, as shown in Fig \ref{fig_principle}a. In this case, the multi-dimensional coding layer consists of a broadband multispectral filter array (BMSFA) and a linear polarization filter array (LPFA). 
We used broadband spectral modulation materials, together with photolithography techniques, to fabricate the BMSFA for spectral modulation \cite{bian2024broadband}. The BMSFA utilizes 16 broadband spectral modulation materials, which combines photoresists with organic dyes for compaction and integration. The materials were arranged in a 4×4 titled array and solidified onto a wafer, enabling pixelation and array formation. 
For polarization encoding, the LPFA is fabricated using nanoimprint lithography, with each spatial pixel corresponding to a miniature linear polarizer. The LPFA consists of 2×2 polarization units, which employ metal wire grids to generate linear polarization states of 0°, 45°, 90°, and 135°. These polarization units are arranged in a periodic mosaic pattern to achieve pixel-level polarization encoding.

\begin{figure}
    \centering
    \includegraphics[width=1\linewidth]{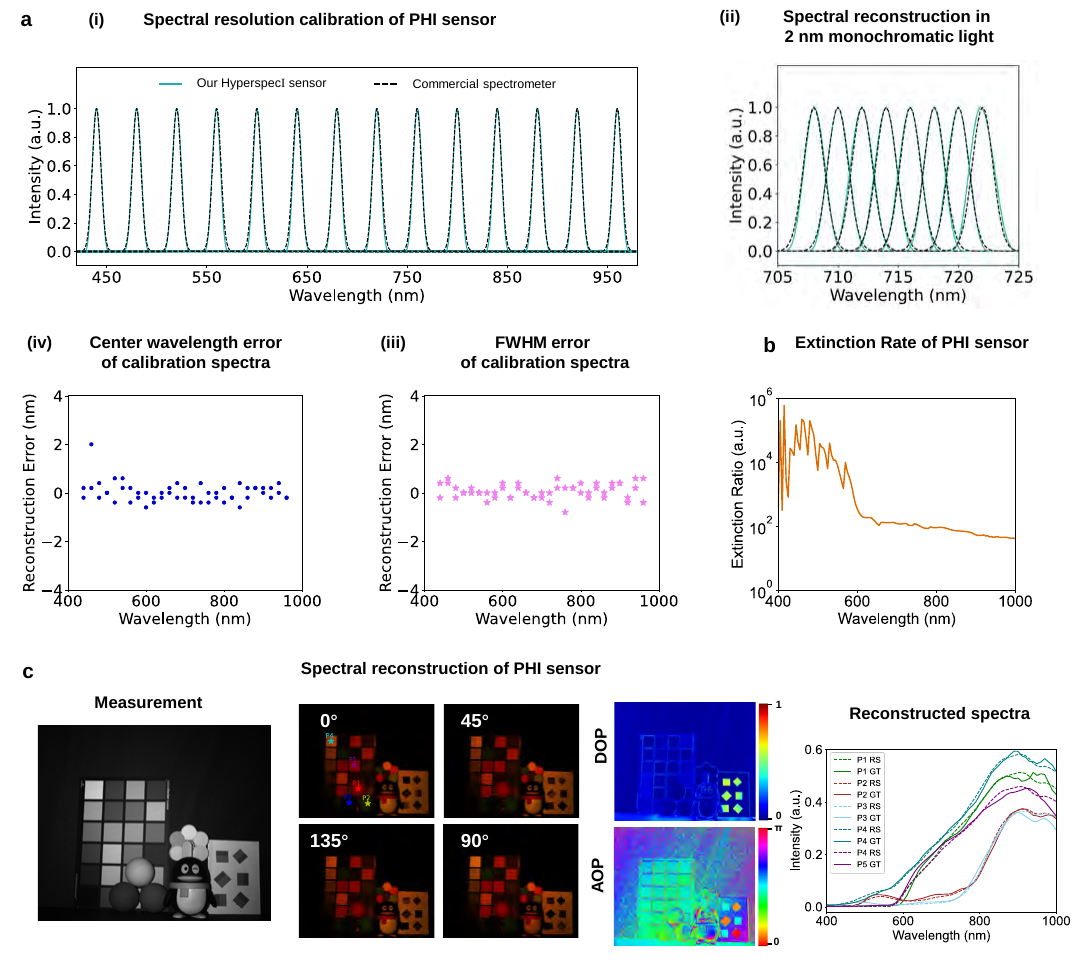}
    \caption{\label{fig_performance} {\textbf{The performance of the PHI sensor.}}
        {\textbf{a.}} (i) Spectral comparison of the PHI sensor (green solid lines) and commercial spectrometer (black dashed lines, Ocean Optics USB 2000+). The calibrated monochrome light was generated by the commercial grating monochromator (Omno151) with 10 nm FWHM. (ii) The results of the extreme spectral resolution using monochrome light with 2 nm FWHM at 0.2 nm spectral interval. (iii-iv) The error of the FWHM and center wavelength between the reconstructed monochrome spectrum and the ground truth.
        {\textbf{b.}} Extinction Rate of PHI sensor in the 400 nm - 1000 nm spectral range.
        {\textbf{c.}} The exemplar hyperspectral and polarization joint imaging results. The 2D measurement is captured by the PHI sensor, containing coupled hyperspectral and polarization information. The reconstructed results of polarization-hyperspectral information under four states are synthesized as RGB images for visualization. Reconstructed spectral images under 0° polarization state in exemplar channels are shown on the right. 
        \\}	
\end{figure}

We conducted both qualitative and quantitative evaluations of the PHI sensor. As shown in Fig. \ref{fig_principle}b, the measurements captured by the PHI sensor contain both polarization and spectral information, encoded by BMSFA and LPFA simultaneously, with 2048×2448 spatial pixels. The high-dimensional hyperspectral and polarization information can be reconstructed using a neural network (PSRNet) at a video rate (see Supplementary Section \ref{supsec_deepLearning} for more details). 
The reconstructed polarization-hyperspectral images have dimensions of 4×61×2048×2448, including hyperspectral information for four linear polarization states (0°, 45°, 90°, and 135°). For each polarization state, the reconstructed HSIs are sampled at 10 nm intervals, resulting in 61 spectral channels.

We utilized the PHI sensor to examine the coupling between polarization and spectral information across different wavelengths, as shown in Fig. \ref{fig_principle}a–b. A white light source was first passed through a linear polarizer to produce linearly polarized light, which then illuminated a plastic sheet exhibiting wavelength-dependent phase retardation. Because of the wavelength-dependent refractive index of the material, it is evident that different wavelengths exhibit distinct polarization characteristics. As illustrated in Fig. \ref{fig_principle}c, the polarization states differ markedly across wavelengths, further demonstrating the system’s effectiveness in jointly sensing spectral and polarization information.

To evaluate the sensor's performance, we employed a monochromator (Omno151) to produce a series of monochromatic light with a 10 nm full width at half maximum (FWHM), which were for calibrating the polarization-hyperspectral sensing matrix of the PHI sensor across different polarization states, as illustrated in Fig. \ref{fig_performance}a(i). Using the calibrated sensing matrix and compressive sensing algorithms, we characterized the sensor's multi-dimensional encoding properties and achieved accurate reconstruction of monochromatic light. For a quantitative analysis of spectral reconstruction accuracy, we compared the FWHM and central wavelength of monochromatic light between the reconstructed results and the commercial spectrometer. The average central wavelength error was 0.22 nm, and the average FWHM error was 0.26 nm, as illustrated in Fig. \ref{fig_performance}a(iii) and Fig. \ref{fig_performance}a(iv). To further validate the system's reconstruction capability under extreme conditions, we tested it with monochromatic light at a 2 nm wavelength, with the reconstructed spectrum presented in Fig. \ref{fig_performance}a(ii).

Additionally, we measured the polarization extinction ratio at different wavelengths to evaluate the PHI sensor's polarization properties across various spectral bands, as presented in Fig. \ref{fig_performance}c. The visible range exhibits a higher polarization extinction ratio, while the extinction ratio in the infrared region is relatively low.
As shown in Fig. \ref{fig_performance}c, the HSIs reconstructed from the four polarization states were synthesized into a pseudo color image for display. By applying the sensor's quantum efficiency (QE) curve, all the reconstructed spectral images were merged into a single-channel polarization degree (DOP) and polarization angle (AOP) image, providing the overall polarization features. We also compared the reconstructed spectra with the ground truth obtained by a commercial spectrometer, achieving an average spectral fidelity of 99.88\%. In addition, we analyze the thermal stability and spatial resolution of our PHI sensor, as detailed in the Supplementary Section \ref{supsec_performance}.

\begin{figurehere}
    \centering
    \includegraphics[width=1\linewidth]{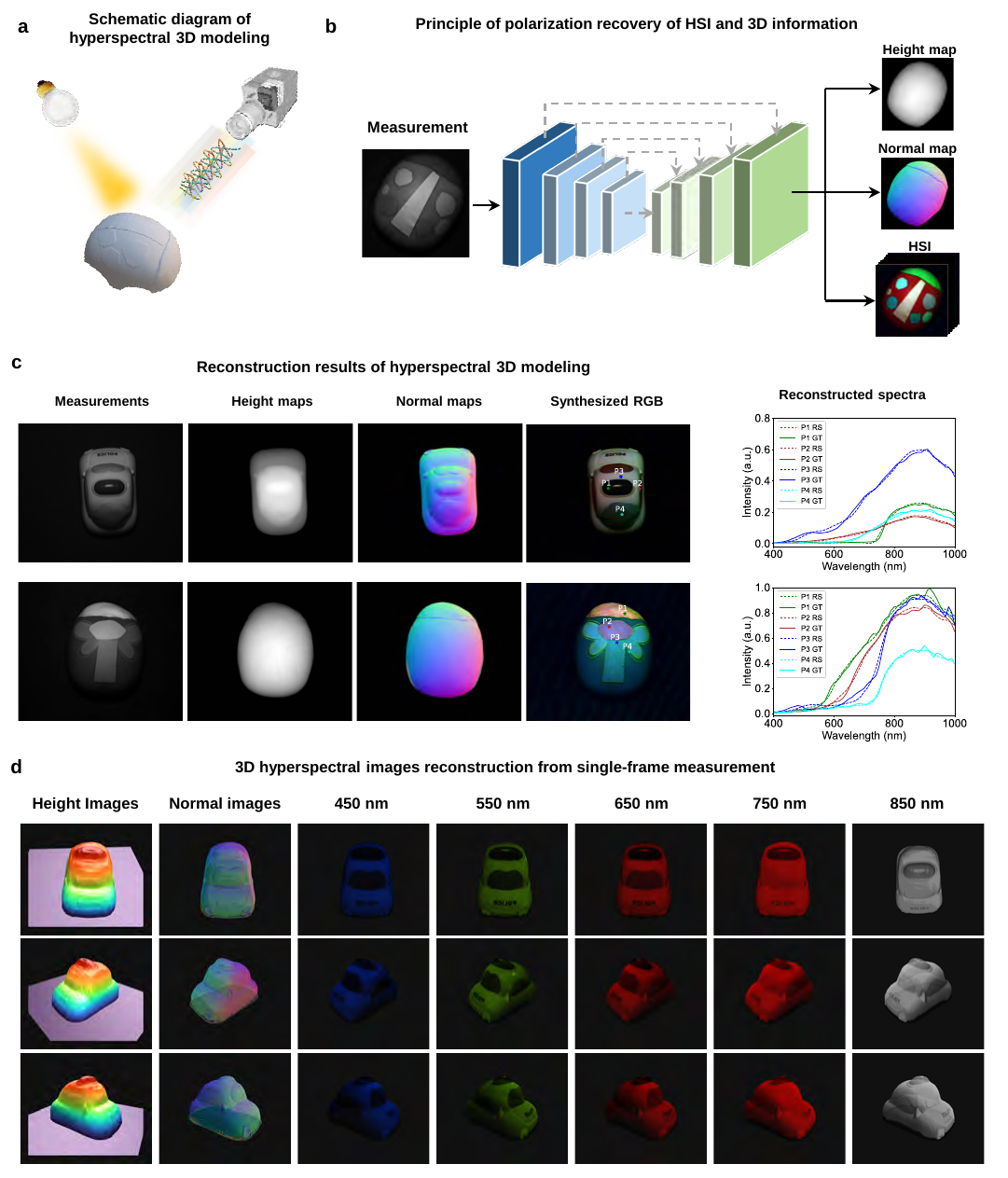}
    \caption{\label{fig_3DModeling}{\textbf{Application for hyperspectral 3D modeling.}}
    {\textbf{a.}} The prototype of hyperspectral 3D modeling. The source illuminates the objects, and the PHI sensor acquires the reflected light.
    {\textbf{b.}} Principle of polarization recovery of HSI and 3D information. The measurements captured by the PHI sensor are processed through an end-to-end restoration network based on UNet framework to obtain the high-dimensional information, including height maps, normal maps, and hyperspectral image (HSI). 
    {\textbf{c.}} Reconstruction results of hyperspectral 3D modeling. The two sample results are displayed. For clarity visualization, height maps are shown in grayscale images, normal maps synthesize surface normal vectors along the [x, y, z] axes as three-channel RGB color images, and the reconstructed HSIs also is synthesized as RGB images. On the right is the comparison between reconstructed spectra (RS) and ground truth (GT) spectra, with the randomly selected points marked in the synthesized RGB images.
    {\textbf{d.}} 3D hyperspectral images reconstruction from single-frame measurement. To better visualize the 3D modeling performance, we present views from three different angles of the results, including 3D height maps, 3D normal maps, and example 3D spectral images at 450 nm, 550 nm, 650 nm, 750 nm, and 850 nm.
    \\}
\end{figurehere}

\section{Application for hyperspectral 3D modeling}\label{sec_3D}



Polarization has been widely applied for geometric feature extraction in 3D reconstruction, where polarized LiDAR enables distance field and surface normal acquisition \cite{scheuble2024polarization}, and multi-view polarization techniques recover 3D morphology through polarized light-field analysis \cite{zhao2022polarimetric,huang2023polarization}.
However, the existing techniques exhibit overemphasis on uni-dimensional geometric analysis with deficient material characterization capabilities. Originating from the simultaneous hyperspectral and polarization imaging abilities of the PHI sensor, here we propose the hyperspectral 3D modeling application, which synergizes hyperspectral imaging with 3D spatial reconstruction to achieve synchronized characterization of high-dimensional spectral and spatial attributes. It provides a novel solution for simultaneous materialogy and morphology analysis.


Specifically, we developed a prototype (Fig. \ref{fig_3DModeling}a) to synchronously capture polarization and hyperspectral data, and prepared a variety of smooth and low-texture objects with minimal surface features. We used our PHI sensor and a 3D scanner (EINSTAR VEGA) to establish a dedicated paired image dataset, containing both polarized spectral data and ground-truth 3D models. Leveraging a deep learning-based approach for polarized hyperspectral 3D reconstruction (Fig. \ref{fig_3DModeling}b), we extracted normal maps, height maps, and HSIs from coupled measurements. Figure \ref{fig_3DModeling}c presents PHI sensor's raw measurements, reconstructed height and normal maps, and the HSIs synthesized in RGB format. The reconstructed spectra, referenced by spatial addresses in the synthesized RGB images, were compared with the ground-truth data captured by the commercial spectrometer, achieving a spectral fidelity of above 99\%. 
Additionally, we compared the reconstructed 3D maps with ground truth from the 3D scanner, producing a PSNR of 35.07 dB for the height maps and 35.32 dB for the normal maps.
As shown in Fig. \ref{fig_3DModeling}d, the reconstruction results are visualized from three different viewpoints, including height maps, normal maps, and spectral images at several example wavelengths (450 nm, 550 nm, 650 nm, 750 nm, 850 nm). Both the quantitative and qualitative experiment results demonstrate the PHI sensor's effectiveness for real-time and high-resolution hyperspectral 3D modeling. More experiment details are referred to the Supplementary Section \ref{supsec_3DModeling}.


\section{Application for hyperspectral sensing against strong reflection and glare}\label{sec_reflectionHDR}

Conventional image sensors suffer significant degradation of imaging quality under complex light conditions, such as specular reflections and strong glare interference, reducing target object identification precision.
Traditional enhancement techniques typically rely on computer vision algorithms to enhance degraded images, such as the removal of reflections or high-dynamic-range (HDR) imaging \cite{zhu2024revisiting}. On the other hand, there are several methods that rely on complex optical elements, such as adaptive optics and optical metamaterials, to address the complex light conditions \cite{metzler2020deep,xiong2023perovskite}. However, the adaptive optics technique builds on the feedback mechanism that is hard for synchronized enhancement, and requires bulky mechanical elements with extremely high cost \cite{vinegoni2016real}. The optical metamaterial technique is limited to laboratory settings and demonstrations, without outdoor experiments for practical applications \cite{xiong2023perovskite}.
Above all, the mentioned techniques are limited to processing monochrome or color images, and do not address the challenges of handling high-dimensional HSIs.

\begin{figurehere}
    \centering
    \includegraphics[width=1\linewidth]{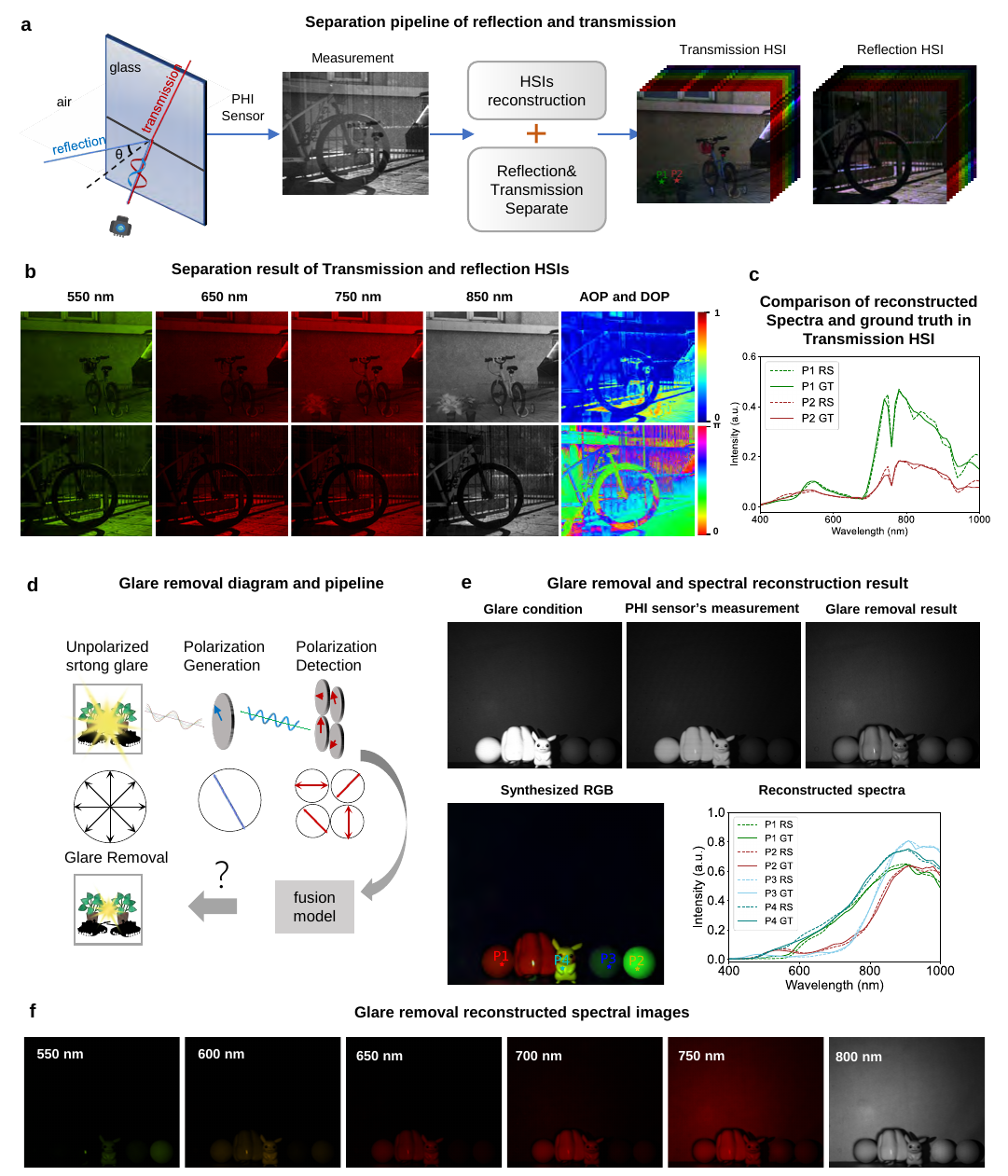}
    \caption{\label{fig_reflection}{\textbf{Application for hyperspectral sensing against strong reflection and glare.}}
    {\textbf{a.}} Pipeline of reflection and transmission HSI separation. The PHI sensor captures measurements through glass, where the acquired data contains aliased spatial-spectral information from both the reflected scene on the same side as the camera and the transmitted scene on the opposite side. Through HSI reconstruction and reflection \& transmission separation, the hyperspectral information for both the transmitted and reflected scenes can be obtained clearly.
    {\textbf{b.}} Separation result of Transmission and reflection HSIs. The example spectral images at 55 nm, 650 nm, 750 nm, and 850 nm are shown on the left. The degree of polarization and angle of polarization are displayed on the right. 
    {\textbf{c.}} Comparison of reconstructed spectra (RS) and ground truth (GT) in Transmission HSI. The selected points of the spectra are marked at transmission HSI in figure a, highlighting the reconstructed accuracy and recognition ability from the differences in spectral dimension between real and artificial plastic plants with the same color. 
    {\textbf{d.}} Glare removal diagram and pipeline. 
    The scene with strong glare is first polarized using a linear polarizer and then captured by the PHI sensor for polarization detection. A fusion model is applied to obtain high-dynamic-range HSIs with glare removal. 
    {\textbf{e.}} Glare removal and spectral reconstruction result. The first row, from left to right, shows the measurement without polarization processing under glare conditions, the original values captured by the PHI sensor, and the glare removal results after fusion processing. On the left of the second row is the reconstructed HDR HSI, which is synthesized as RGB images for visualization. On the right, the performance of the reconstructed spectra is compared with the ground truths contained by the commercial spectrometer. 
    {\textbf{f.}} Reconstructed spectral images for glare removal at exemplar wavelengths.\\}
\end{figurehere}


We applied the PHI sensor to exploit polarization-spectral information under complex light conditions, and achieved accurate material characterization and anti-counterfeiting identification. We used the PHI sensor to acquire images corrupted by nonuniform illumination by leveraging the inherent refractive index contrast in multilayer dielectric media (glass/air interfaces). By physics-driven data modeling, we developed a separation strategy to disentangle reflection and transmission components from measurements, as shown in Fig. \ref{fig_reflection}a. Subsequently, we performed a spectral reconstruction for both decomposed scenes, visualizing the reconstructed HSI in the RGB format and validating it against commercial spectrometer measurements (Fig. \ref{fig_reflection}b).
As illustrated in Fig. \ref{fig_reflection}c, the reconstructed spectra of real (P1) and artificial plastic (P2) green plants in the transmitted scene show accurate detection of metamerism, a phenomenon where objects appear similar in color but have different spectra. More experiment and result details are referred to the Supplementary Section \ref{supsec_applicationReflection}.

Glare interference significantly degrades visual quality, hindering accurate hyperspectral imaging. To address this challenge, our PHI sensor employs an innovative polarization modulation technique that effectively suppresses glare, enabling the acquisition of HDR HSI in a single frame exposure. By combining the PHI sensor with a linear polarizer, each frame records photoelectric responses under four orthogonal polarization states (0°, 45°, 90°, and 135°), obtaining coupled measurements with varying texture and spectral details. These images under four states are then adaptively fused using our proposed HDR algorithm, which incorporates exposure suppression and linear weighted fusion to enhance dynamic range while preserving spectral fidelity and eliminating glare (Fig. \ref{fig_reflection}d).
As shown in Fig. \ref{fig_reflection}e, glare-affected regions in the raw data exhibit significant overexposure, obscuring fine spatial and spectral details. However, the measurement of the PHI sensor contains more detailed information under four polarization states, which can improve the HDR of reconstruction HSI and achieve glare removal. The results of glare removal hyperspectral imaging are shown in the synthesized RGB image and displayed using spectral images in Fig. \ref{fig_reflection}f.
A comparison of the reconstructed spectral curves with ground truth measurements obtained by a commercial spectrometer demonstrates a spectral fidelity of 99.93\%. These results confirm the PHI sensor’s capability for high-precision spectral sensing under complex light conditions (see the Supplementary Section \ref{supsec_applicationHDR} for more details).





\section{Conclusion and Discussion}\label{sec_conclusion}

This work reports a multi-dimensional on-chip optical imaging (MOCI) architecture that utilizes optical encoding and computational imaging theories to simultaneously acquire multiple high-dimensional optical information. The MOCI consists of three key components, including the multi-dimensional encoding layer, the image acquisition layer, and the computational reconstruction layer, which together facilitate the encoding, acquisition, and decoding of multi-dimensional optical data. Building on this architecture, we successfully fabricated a polarization-hyperspectral imaging (PHI) sensor that integrates a broadband multispectral filter array and a linear polarization filter array, achieving joint spectral-polarization encoding using a single sensor. The sensor acquires the original multi-dimensional optical information at video rates (74 FPS) with high spatial-spectral-polarization resolution (2048 $\times$ 2448 pixels $\times$ 61 channels $\times$ 4 angles), which is reconstructed from the coupled measurements using a deep-learning neural network. 
The experiments demonstrate that the PHI sensor produces advanced performance under complex light conditions, enabling hyperspectral separation of reflected and transmitted scenes and high-dynamic-range hyperspectral imaging with glare removal. Additionally, the sensor achieves hyperspectral 3D modeling, leveraging polarized structured information to obtain height maps, normal maps, and point cloud data from a single frame of coupled measurement.
Compared to other snapshot computational imaging techniques that are typically designed to capture only specific optical dimensions, the MOCI technique offers advantages including multiple imaging dimensions, high integration, high resolution, and high video rate, enabling snapshot acquisition of compressive optical information and providing a more comprehensive characterization of target properties, potentially driving the advancement of next-generation intelligent imaging and sensing devices.

The successful demonstration of the PHI sensor highlights the effective MOCI architecture, which can be further extended for advanced multi-dimensional optical sensing.
First, in the view of compressive encoding, advanced optical encoding structures \cite{miao2025computational,li2023imaging}, optical coding materials \cite{li2024imaging,wang2016optically}, and micro-electromechanical systems (MEMS) \cite{meng2021dynamic,arbabi2018mems} can be introduced to further acquire phase, light field, and other high-dimensional optical information within a single miniature sensor.
Second, in view of computational decoding, advanced deep learning, and big model techniques can be introduced to further enhance reconstruction and unmixing precision, helping dig into the multi-dimensional optical correlation \cite{lecun2015deep}.
Third, the above computation can be packed into a dedicated chip, which can further help integration without requiring additional high-power computational graphics \cite{rogers2021universal,chen2023all}.

\newgeometry{left=1.2cm, right=1.2cm, top=2.0cm, bottom=2.0cm}
\begin{table}[!h]
    \renewcommand{\arraystretch}{2.5} 
    \begin{center}
        \caption{Comparison of different polarization-spectral sensing techniques.}\label{Table_compare}
        \setlength{\tabcolsep}{1 pt}
        \footnotesize
       
        \begin{tabular}{m{0.8in} <{\centering}m{0.6in}<{\centering} m{.6in}<{\centering} m{.7in}<{\centering} m{0.7in}<{\centering} m{0.7in}<\centering m{0.8in}<\centering }
            \toprule
            \textbf{Integration type} & \textbf{Imaging mode} & \textbf{Spatial resolution (pixel)} & \textbf{Polarization channel} & \textbf{Spectral range } & \textbf{Spectral resolution}&\textbf{Ref}  \\
            \midrule
           
            \cellcolor{lightgray!25}&\cellcolor{lightgray!20}  & \cellcolor{lightgray!10}1024 × 1024 &\cellcolor{lightgray!10} Full-Stokes&\cellcolor{lightgray!10}400-1000 nm &\cellcolor{lightgray!10}5 nm &\cellcolor{lightgray!10}Opt. Express 2021 \cite{bai2021static}\\ 

              \cellcolor{lightgray!25}&\cellcolor{lightgray!20} 
               & 1 × 1 &  Full-Stokes & 1400-1500nm & 2nm &eLight 2022 \cite{ni2022computational}  \\
              
              \cellcolor{lightgray!25}&\cellcolor{lightgray!20}  &\cellcolor{lightgray!10}  640 × 480 &\cellcolor{lightgray!10}  4 channels (0$^\circ$, 45$^\circ$, 90$^\circ$, 135$^\circ$)&\cellcolor{lightgray!10} 8-14$\mu$m  &\cellcolor{lightgray!10}0.1$\mu$m&\cellcolor{lightgray!10}Optica 2024 \cite{wang2024spinning}\\  
            
            \cellcolor{lightgray!25}&\cellcolor{lightgray!20} &1936 × 1464& 4 channels (0$^\circ$, 45$^\circ$, 90$^\circ$, 135$^\circ$) & 400-750 nm & 25 nm& eLight 2024\cite{wang2024exploiting} \\ 
           
             \cellcolor{lightgray!25}&\cellcolor{lightgray!20}\multirow{-6}{*}{Scanning} &\cellcolor{lightgray!10}  2016 × 2016  & \cellcolor{lightgray!10}Full-Stokes &\cellcolor{lightgray!10} 400-700nm &\cellcolor{lightgray!10} 28 channels&\cellcolor{lightgray!10}Laser Photonics Rev. 2025 \cite{wen2025full}  \\
             
             \cellcolor{lightgray!25}& \cellcolor{lightgray!10}    & 400 × 400& Full-Stokes& 400-850nm & 72 channels &Adv. Photonics Nexus 2023 \cite{han2023deep}\\

             \cellcolor{lightgray!25}& \cellcolor{lightgray!10}&\cellcolor{lightgray!10}  451 × 360 &\cellcolor{lightgray!10} Full-Stokes&\cellcolor{lightgray!10} 400-900 nm &\cellcolor{lightgray!10} 1 nm &\cellcolor{lightgray!10}Nature  2024\cite{fan2024dispersion} \\

             \cellcolor{lightgray!25}\multirow{-10}{*}{System}&   \multirow{-3}{*}{Snapshot}\cellcolor{lightgray!10}&$\textless$640 × 512  & Full-Stokes & 1520-1620 nm & 2.5 nm&Nano Lett. 2024 \cite{chen2024imaging} \\  
             \arrayrulecolor{white}
             \cmidrule(lr){1-1}
             \arrayrulecolor{black}
            
        \cellcolor{lightgray!25}& \cellcolor{lightgray!20}&\cellcolor{lightgray!10} 1 × 1  &\cellcolor{lightgray!10} 3 channels (0$^\circ$, 45$^\circ$, 90$^\circ$)&\cellcolor{lightgray!10} 400-750 nm  &\cellcolor{lightgray!10} 15 channels  &\cellcolor{lightgray!10} Sci. Adv. 2021\cite{altaqui2021mantis}\\

           \cellcolor{lightgray!25}&\cellcolor{lightgray!20}& 1 × 1 & Full-Stokes &  1-8$\mu$m &0.5$\mu$m&Nat.Commun. 2024 \cite{jiang2024metasurface}\\
            
           \cellcolor{lightgray!25}&\multirow{-4}{*}{Scanning}\cellcolor{lightgray!20}&\cellcolor{lightgray!10}   1 × 1  &\cellcolor{lightgray!10} Full-Stokes &\cellcolor{lightgray!10} 1476-1599 nm &\cellcolor{lightgray!10} 10 channels&\cellcolor{lightgray!10}Nat. Photonics 2025 \cite{tang2025adaptive} \\
            
             
              \cellcolor{lightgray!25}&\cellcolor{lightgray!10}&  42 × 50  & Full-Stokes & 700-1150nm & 0.23 nm &Sci. Adv. 2024 \cite{zhang2024real}  \\

               \multirow{-7}*{On-chip}\cellcolor{lightgray!25}&  \multirow{-2}{*}{Snapshot}\cellcolor{lightgray!10}&\cellcolor{lightgray!10}\textbf{2048×2448}  &\cellcolor{lightgray!10}\textbf{4 channels (0$^\circ$, 45$^\circ$, 90$^\circ$, 135$^\circ$)} & \cellcolor{lightgray!10}\textbf{400-1000 nm} &\cellcolor{lightgray!10}\textbf{61 channels}&\cellcolor{lightgray!10}\textbf{MOCI 2025} \\
            \bottomrule
        \end{tabular}
    \end{center}
\end{table}
\restoregeometry

\newpage
\bibliography{reference}




%

\newpage




\newpage

\noindent\textbf{Data availability} All data generated or analyzed during this study are included in this published article and the public repository at the following link: https://github.com/bianlab/MOCI.\\

\noindent\textbf{Code availability} The demo code of this work is available from the public repository at the following link: https://github.com/bianlab/MOCI.\\

\noindent\textbf{Acknowledgements} This work was supported by the National Natural Science Foundation of China (62322502, 62088101, 624B2028). We thank Yuzhe Wang for helping conduct the experiments.\\

\noindent\textbf{Author contributions} L.B. and Z.W. conceived the idea. Z.W. and L.B. designed and fabricated the modulation layers, and implemented sensor integration. Z.W. and P.P. calibrated the sensor and tested its imaging performance. P.P., Z.Z., and Z.W. designed and implemented the hyperspectral 3D reconstruction experiment. Z.W. and P.P. conducted experiments on hyperspectral sensing against strong reflection and glare. L.B., Z.W., P.P., Z.Z., R.Y., H.X., and J.Z. prepared the figures and wrote the manuscript with input from all the authors. L.B. and J.Z supervised the project.\\

\noindent\textbf{Competing interests} L.B., Z.W., and J.Z. hold patents on technologies related to the devices developed in this work (China patent numbers ZL 2024 1 0797667.2, and ZL 2022 1 0764166.5) and submitted related patent applications.\\


%
\newpage

\end{document}